\documentclass{Pos}

\title{Lorentz shift measurements  in heavily irradiated silicon detectors in high magnetic fields}

\ShortTitle{ Lorentz shifts in heavily irradiated silicon sensors }

\author{Wim de Boer$^1$, Karl-Heinz Hoffmann$^1$, Andreas Sabellek$^1$, \speaker{Mike Schmanau}$^1$, Michael Schneider$^1$, Valery Zhukov$^1$ and Theo Schneider$^2$\\
        $^1$Institute of Experimental Nuclear Physics and $^2$Institute of Technical Physics, \\ Karlsruhe Institute of Technology,
        Herrmann-von-Helmholtz-Platz 1, 76344 Karlsruhe, Germany\\
        E-mail: \email{schmanau@cern.ch}}

\abstract{
An external magnetic field exerts a Lorentz force on drifting electric charges inside a silicon strip sensor  and thus shifts the  cluster position of the collected charge.
The shift can be related to the Lorentz angle which is
typically a few degrees for holes and a few tens of degrees for electrons in a
4\,T magnetic field. The Lorentz angle depends upon magnetic field, electric field inside the sensor and
temperature. In this study the sensitivity to radiation for fluences up to 10$^{16}$\,n$_{eq}$/cm$^2$
has been studied.

The Lorentz  shift   has been measured by inducing ionization with 670 nm red or 1070 nm infrared laser
beams injected into the back side of the irradiated silicon sensor operated in magnetic fields up to 8 T.
For holes the shift as a function of radiation is increasing, while for electrons it is decreasing and even changes  sign.
 The fact that for irradiated sensors the Lorentz shift for electrons is smaller than for holes, in contrast to the observations in non-irradiated sensors, can be qualitatively explained by  the structure of the electric field in irradiated sensors.
}

\FullConference{9th International Conference on Large Scale Applications and Radiation Hardness of Semiconductor Detectors,RD09\\
		September 30-October 2, 2009\\
		Florence,Italy}

\begin{document}
\section{Introduction}
% notations
The Lorentz shift $\Delta x$ for the silicon strip detector of thickness D with a magnetic field B perpendicular
to the electric field  E inside the sensor can be written as \cite{smith}:
$\rm  \Delta x= D  tan(\Theta_{\mathrm{L}})=D r_{\mathrm{\mbox{\tiny H}}} \mu B$, where $ r_{\mathrm{\mbox{\tiny H}}}$  is the Hall factor ($\approx\;$0.7 for holes and $\approx\;$1.15 for
electrons at room temperature \cite{lb}) and $\mu$ is the  drift mobility. The drift mobility is known to increase strongly with
decreasing temperature and is proportional to the electric field until
saturation is reached\cite{JacoboniCanali:1977}. The drift mobility is a factor two to three larger for electrons than for holes and together with the different Hall factors $r_{\mathrm{\mbox{\tiny H}}}$ the Lorentz  shift for
electrons is a factor of four larger than the one for holes, at least in non-irradiated sensors.

% different sensors
The radiation is not expected to change the mobility significantly up to 10$^{16}$\,n$_{eq}$/cm$^2$,
but the E-field will because of the space charge being trapped in the defects caused by irradiation.
This space charge leads to strongly non-uniform electric fields with a double peak structure \cite{eremin1}, which
makes it difficult to predict the Lorentz shift, so in this study the shift has been  measured experimentally
using different silicon sensors. The sensors have been irradiated at the proton cyclotron  of KIT, Karlsruhe
with 25 MeV protons and studied using laser beams inside the superconducting JUMBO magnet in the magnet lab of KIT,
Karlsruhe, which features a warm bore \cite{jumbo}.

\section{Experimental setup}
\begin{figure}
\begin{center}
\includegraphics [width=9cm,clip]{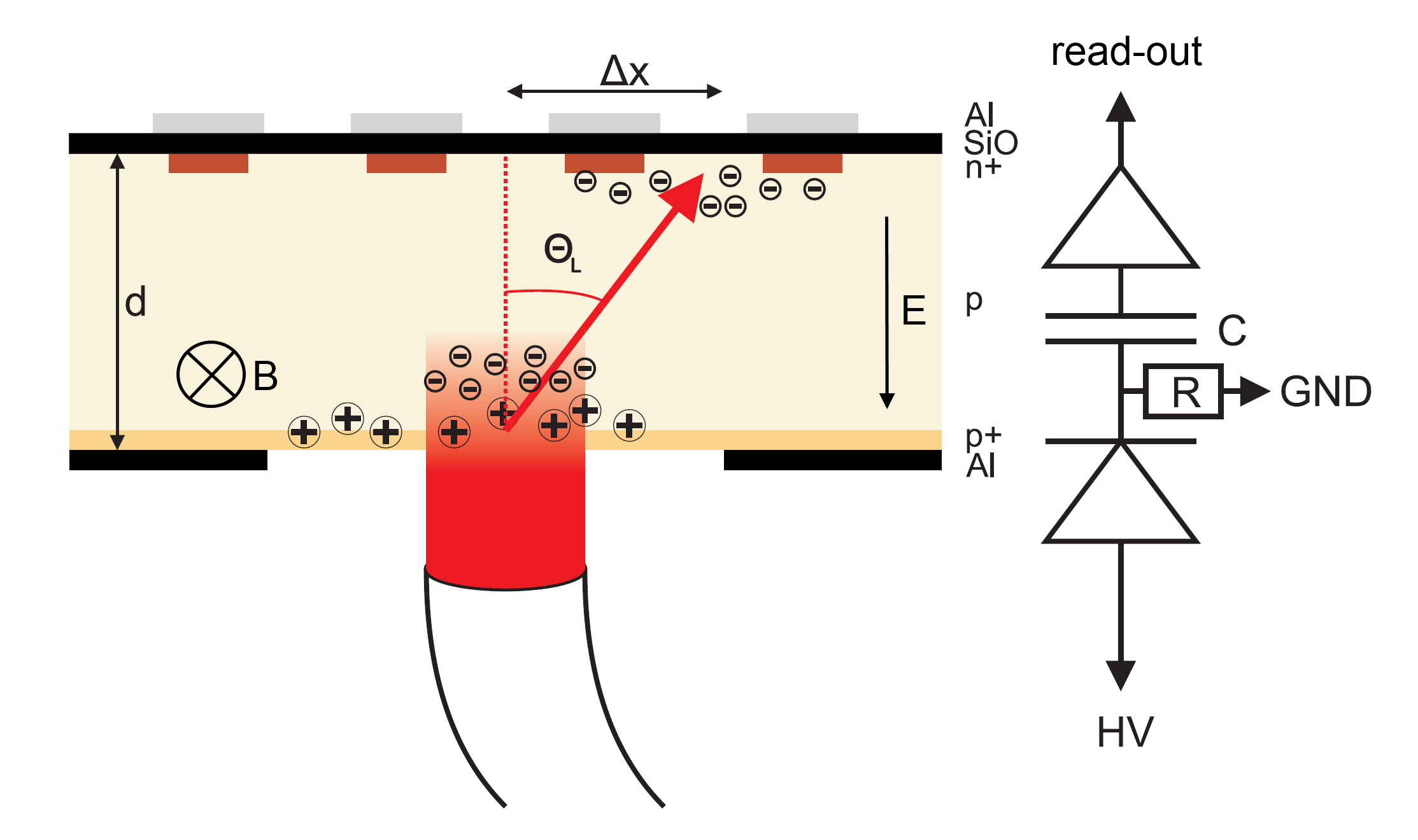}
\caption{\label{f1} \it The principle of the Karlsruhe setup to measure the Lorentz angle of holes and electrons:
charge is locally generated at the backplane by a laser pulse and the Lorentz shift is measured by the shift in
distance compared with the position at zero magnetic field.}
\end{center}
\end{figure}

% setup
The detailed description of the laser setup can be found in several previous publications
\cite{roederer:1998,heising:1999,hauler:2000,osaka:2000,Schneider:2009} and the principle is shown in Fig.\,\ref{f1}.
Two lasers have been used:  a) the red one with  a wavelength
of 670 nm\footnote{The laser QFLD-670-2S from QPhotonics} penetrates only a few $\mu$m
inside the backplane of the sensors, b) the infrared  1060 nm
laser\footnote{LD-1060 from Fermionics Lasertech Inc.} creates a trajectory inside the whole volume.
Both lasers had a maximum power of
1\,mW adjustable by the pulse height and width of the input
pulse to the laser diode. The  laser pulse with a width
of $\sim$1\,ns was generated and transmitted to the sensor via a
single-mode fiber with an inner diameter of a few microns. The beam spot on the detector can be varied from a
few 10 micron onwards by changing the distance between fibre and sensor.
The setup was located inside the JUMBO magnet operating with the B field up to 8\,T.
The temperature of the sensor was controlled and kept nearly constant throughout the measurements
by flux of cold nitrogen.

% electronics
The readout electronics include Premux128-Chip
with a shaping time of 45\,ns \cite{jones} has a built-in
common mode suppression by a double correlated sampling
technique. The analog signals of the 128 Premux channels are multiplexed and
digitized by an external ADC.
The time delays have been adjusted to deliver a maximum signal from the laser.
The high voltages and leakage currents were controlled by a Keithley 2410
with a maximum voltage of 1100\,V. The cabling limits the studied range of depletion
voltages to 1000\,V.

\begin{table}[]
\caption[]{\label{sensortable} \it List of sensors with  strips on one side (the p-side (n-side) for n-type (p-type) bulk material) and their parameters. The sensors were irradiated at the
proton cyclotron of KIT, Karlsruhe with different fluences $\Phi$ (25\,MeV Protons, hardnessfactor 1.85).
The depletion voltage U$_{dep}$ was derived from the knee in the 1/C over U plot.
}\vspace*{2mm}
\begin{tabular}{|c|c|c|c|c|c|c|}
\hline
Sensorname & Manufacturer & Material & d [${\mu}m$] & Pitch [${\mu}m$] & U$_{dep}$ & $\Phi$ [n$_{eq}$/cm$^2$] \\
\hline
\hline
CMS01 Mini & ST & FZ n-Type & 500 & 120 & 154 & 0  \\
\hline
RD50 28-3-1  & Micron & FZ p-Type & 300 & 80 & 12 & 0\\
RD50 27-3-2  & Micron & FZ p-Type & 300 & 80 & $\sim$1000 & 1$\cdot$10$^{15}$\\
RD50 28-7-3  & Micron & FZ p-Type & 300 & 80 & $>$1000 &  9.6$\cdot$10$^{15}$\\
\hline
MCz 0802-1 & HIP & MCz n-Type & 300 & 50 & 169 & 7.1$\cdot$10$^{14}$  \\
MCz 0802-5 & HIP & MCz n-Type & 300 & 50 & 272 & 7.1$\cdot$10$^{14}$  \\
MCz 0802-3 & HIP & MCz n-Type & 300 & 50 & $>$1000 & 7.2$\cdot$10$^{15}$  \\
MCz 0802-9 & HIP & MCz n-Type & 300 & 50 & 347 & 0  \\
\hline
\end{tabular}
\end{table}

% sensors tested
Three types of sensors have been measured: Float Zone (FZ) n-type from STMicroelectronics and p-type from Micron, and Chochalski (MCz) n-type from Helsinki Institute of Physics (HIP), see Table\,\ref{sensortable}. The p-type sensors with n-strips have been used for studying the electrons and
the n-type sensors with p-strips for holes, since the charge is integrated on the n-strips for electrons and on the p-strips for holes in reversely biased diodes.
Before irradiation all n-type sensors were fully depleted at a
bias voltage of $\leq$347\,V, the p-type sensor already at 12\,V. After irradiation with $\Phi>$1.0$\cdot$10$^{15}$\,n$_{eq}$/cm$^2$ all depletion voltages
exceed 1000\,V and a full depletion was not possible.

\section{Results}
\begin{figure}
\begin{center}
\includegraphics [width=0.35\textwidth,clip,angle=90] {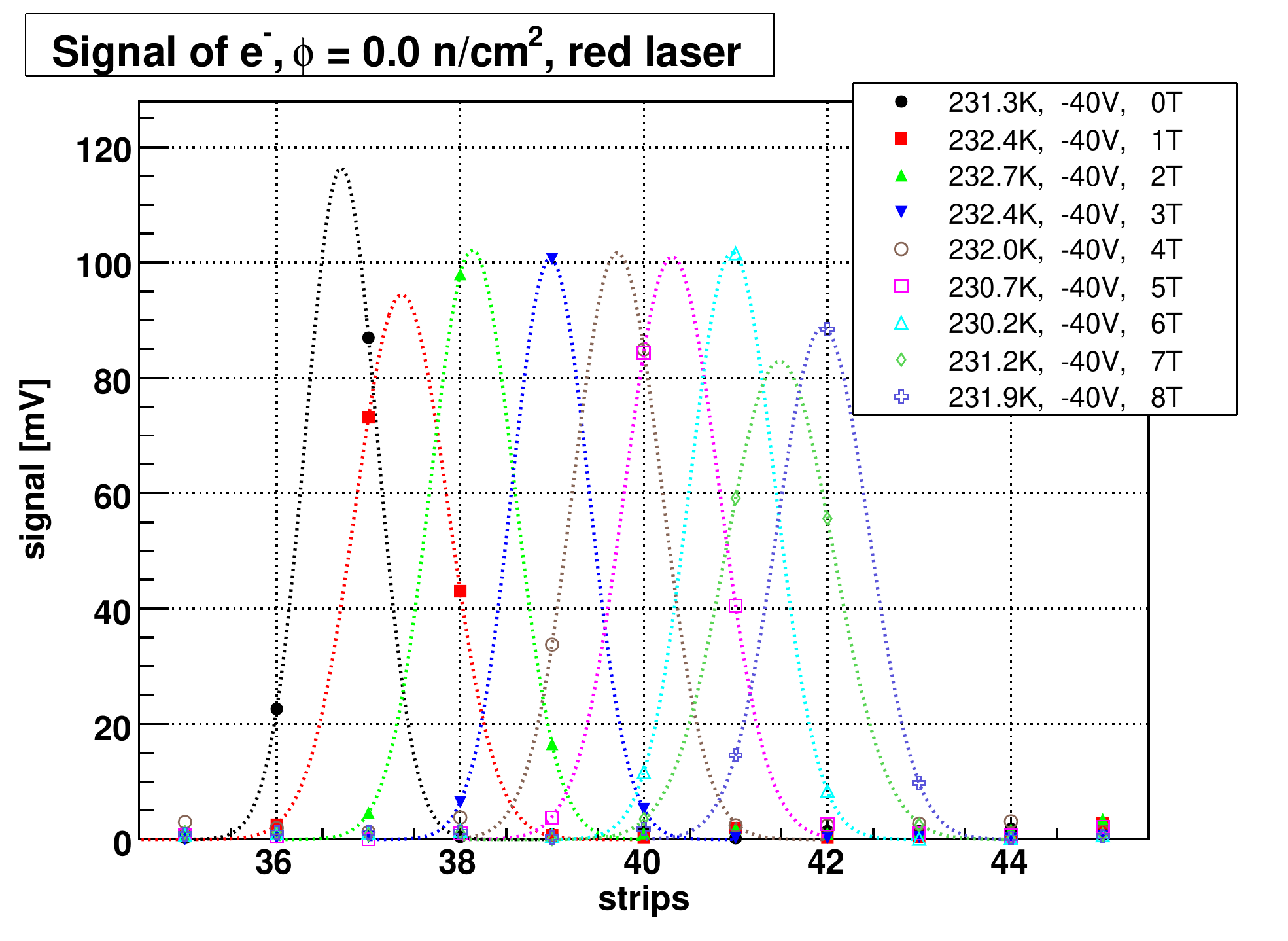}
\includegraphics [width=0.35\textwidth,clip,angle=90] {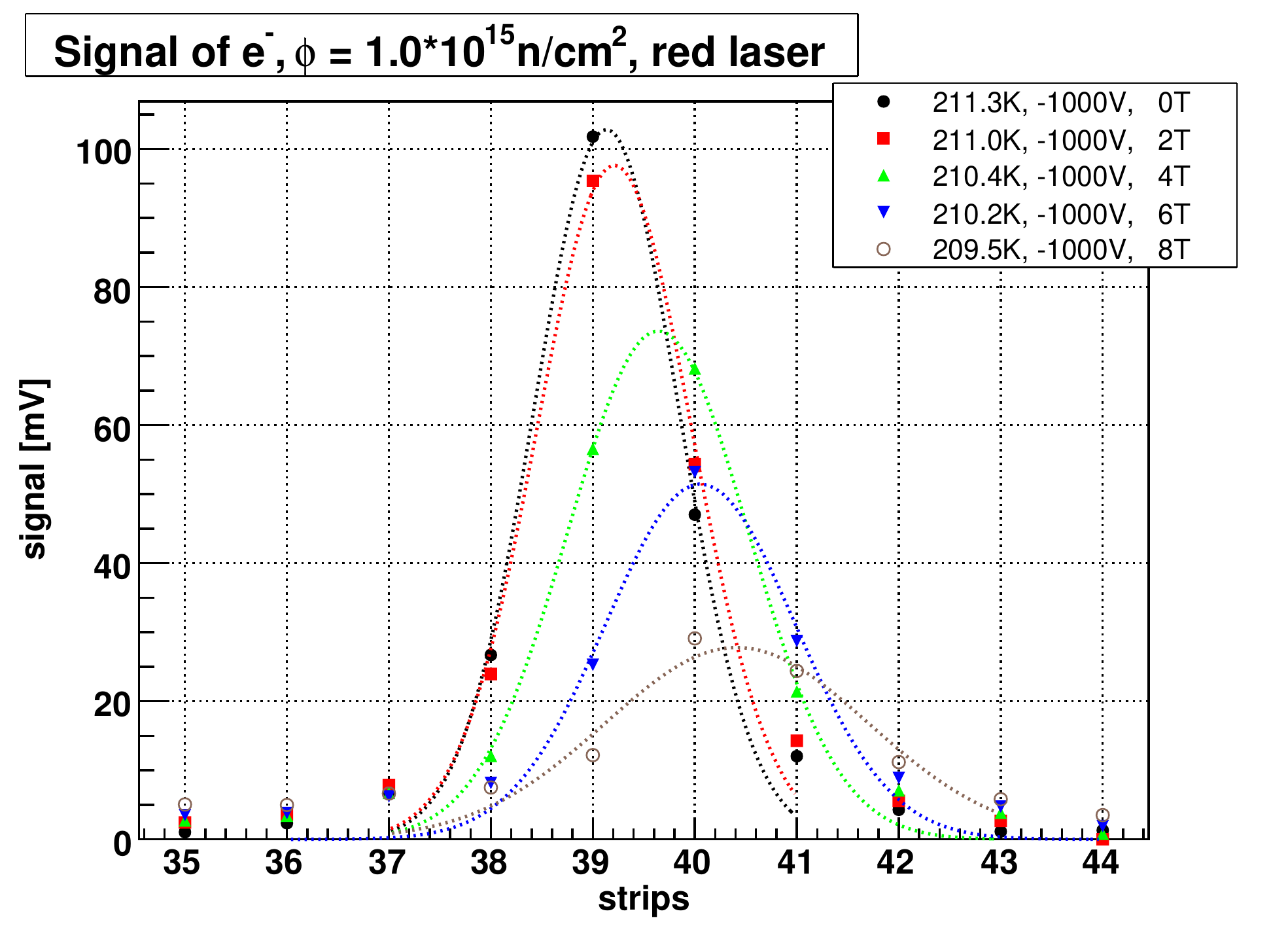}
\caption{\label{f2} \it Left:
the signal for electrons in a non-irradiated sensor using a red laser. Right: as left for a fluence
of $\phi = 1.0 * 10^{15}$\,$n_{eq}$/cm$^2$. One observes the strong reduction in  Lorentz shift, which is largely due to the higher
bias voltage needed to reach depletion.  Note that the shift is in the same direction for both, the unirradiated and irradiated sensor, i.e. it has the same sign.}
\end{center}
\end{figure}

\begin{figure}
\begin{center}
\includegraphics [width=0.35\textwidth,clip,angle=90] {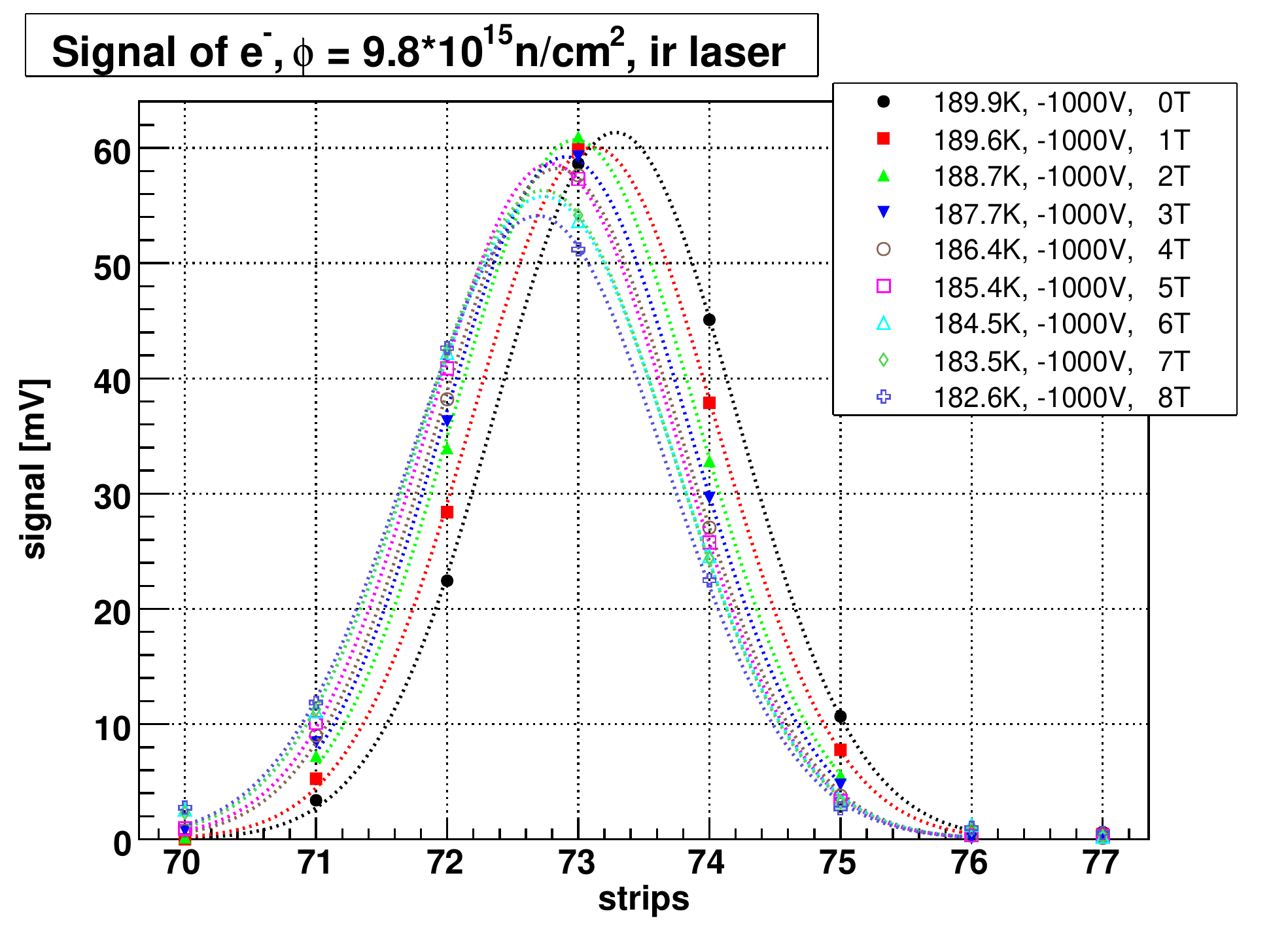}
\includegraphics [width=0.35\textwidth,clip,angle=90] {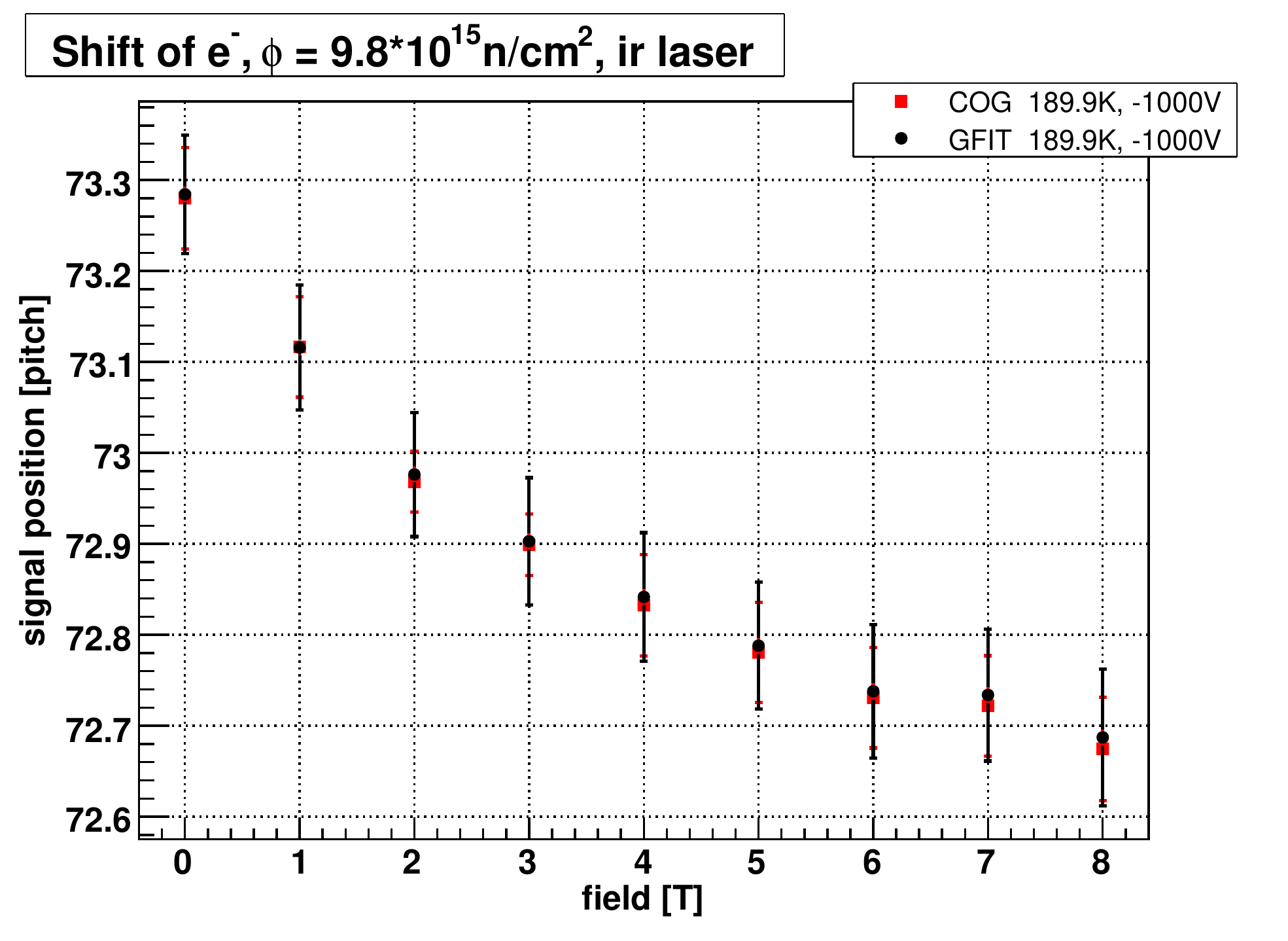}
\caption[]{\label{f3} \it
 The raw n+ strip signals (collecting electrons) in different magnetic fields with the fitted gaussian curves (left) and the peak position of the fitted gaussian curve (GFIT) or center of gravity (COG) as function of B-field (right). The  sensor was irradiated with a fluence of about $10^{16}$\,$n_{eq}$/cm$^2$ and the signals were generated with the infrared laser. The shift is given in units of the pitch between the strips, which is 80 $\mu m$.}
\end{center}
\end{figure}

\begin{figure}
\begin{center}
\includegraphics [width=0.35\textwidth,clip,angle=90] {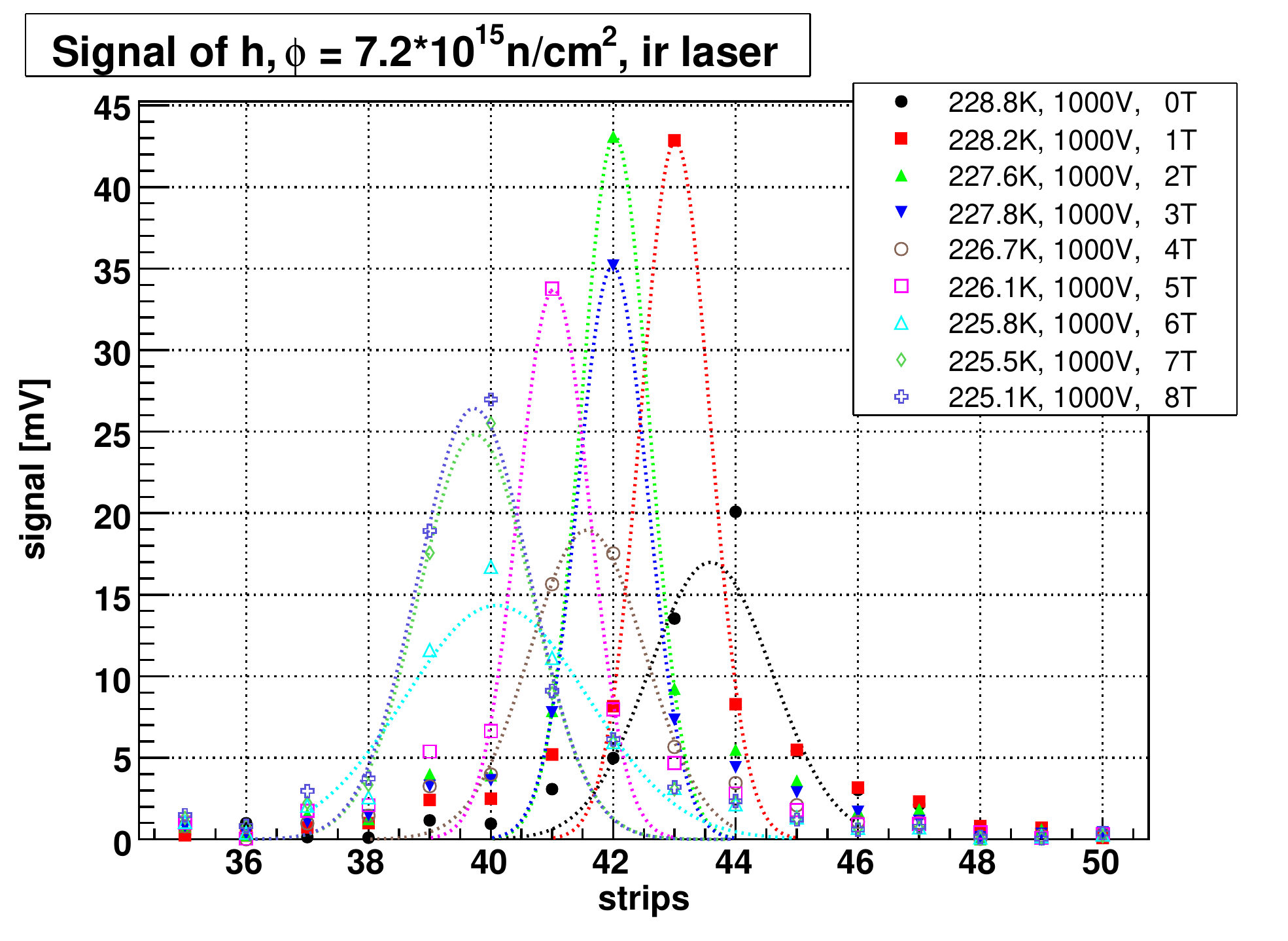}
\includegraphics [width=0.35\textwidth,clip,angle=90] {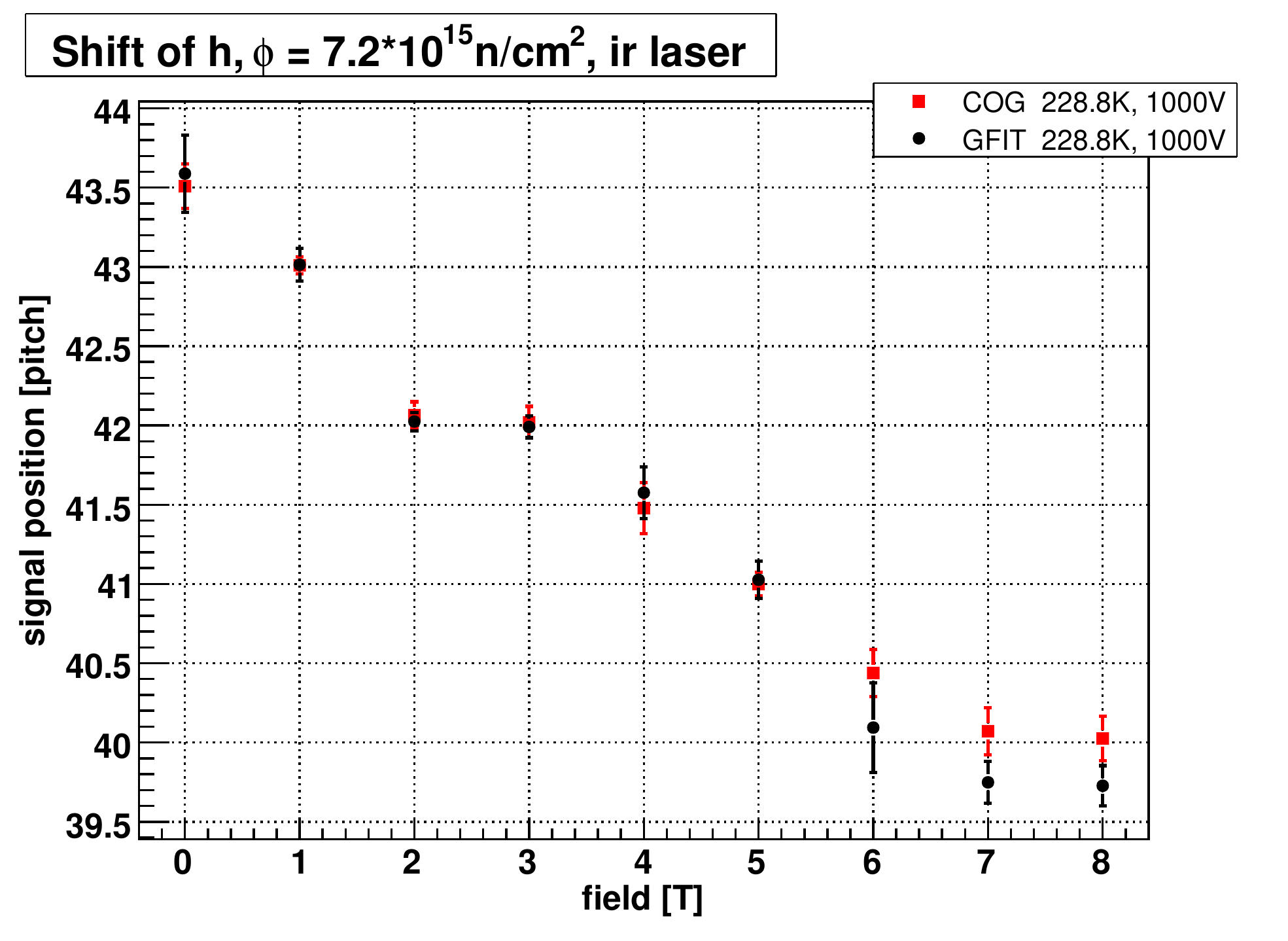}
\caption[]{\label{f4} \it
 The raw p+ strip signals (collecting holes) in different magnetic fields with the fitted gaussian curves (left) and the peak position of either the fitted gaussian curve (GFIT) or center of gravity (COG) as function of B-field (right). The  sensor was irradiated with a fluence of about $10^{16}$\,$n_{eq}$/cm$^2$ and the signals were generated with the infrared laser. The shift is given in units of the pitch between the strips, which is 50 $\mu m$.}
\end{center}
\end{figure}

% signal profiles
The signals for different sensors and  B-fields are shown in Figs.\,\ref{f2},
\ref{f3} and \ref{f4}. The sensor temperatures (around 200\,K) have been more accurately indicated in the figures. The width of the signals is determined by the width of the laser beam.
The Lorentz shift of the signal with increasing B-field is clearly visible.
The  measurements  confirm that the  Lorentz shift increases linearly with the
magnetic field up to 8\,T \cite{osaka:2000}.

\begin{figure}
\begin{center}
\hspace {0mm}
\includegraphics [width=0.50\textwidth,clip] {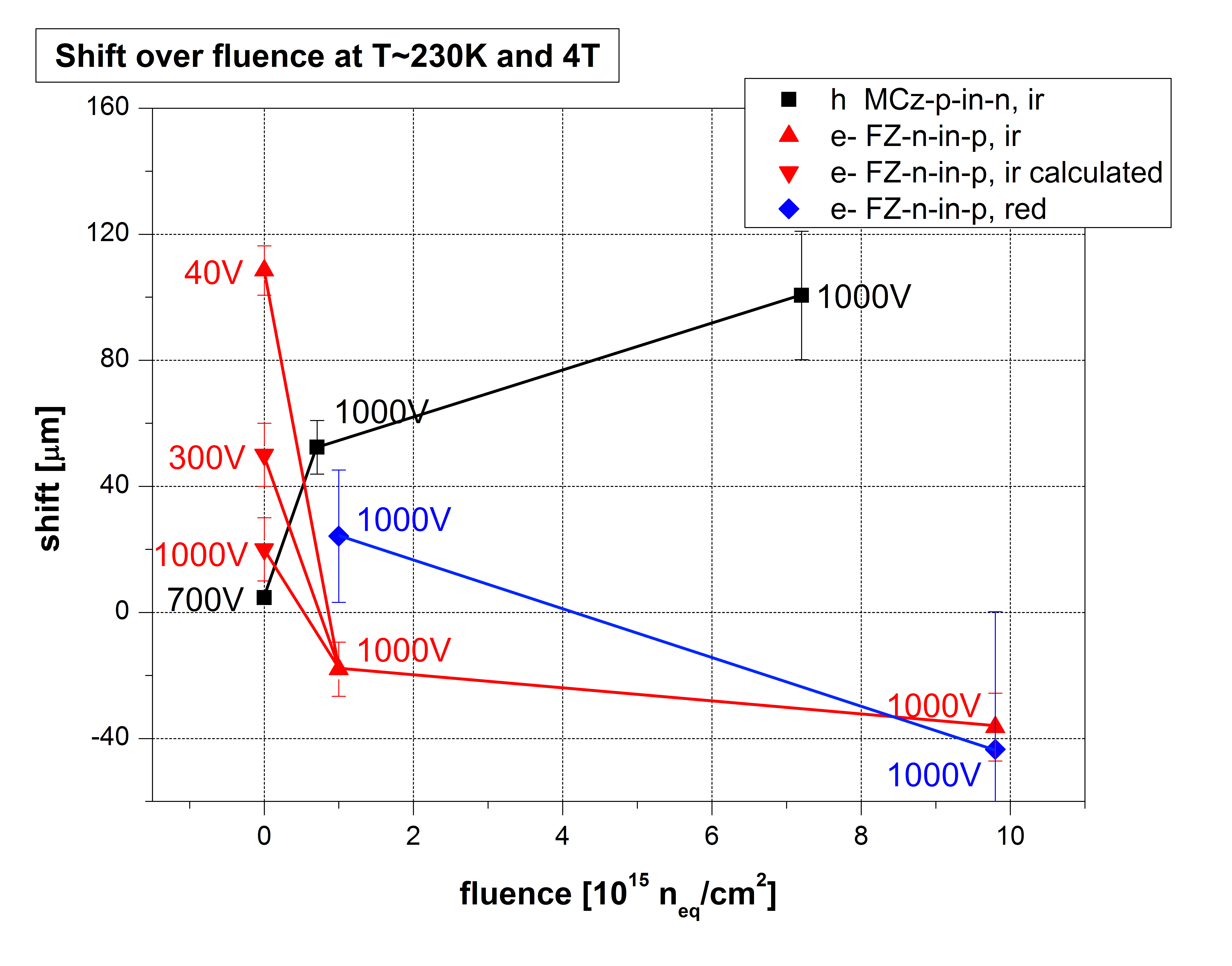}
\includegraphics [width=0.47\textwidth,height=0.38\textwidth,clip] {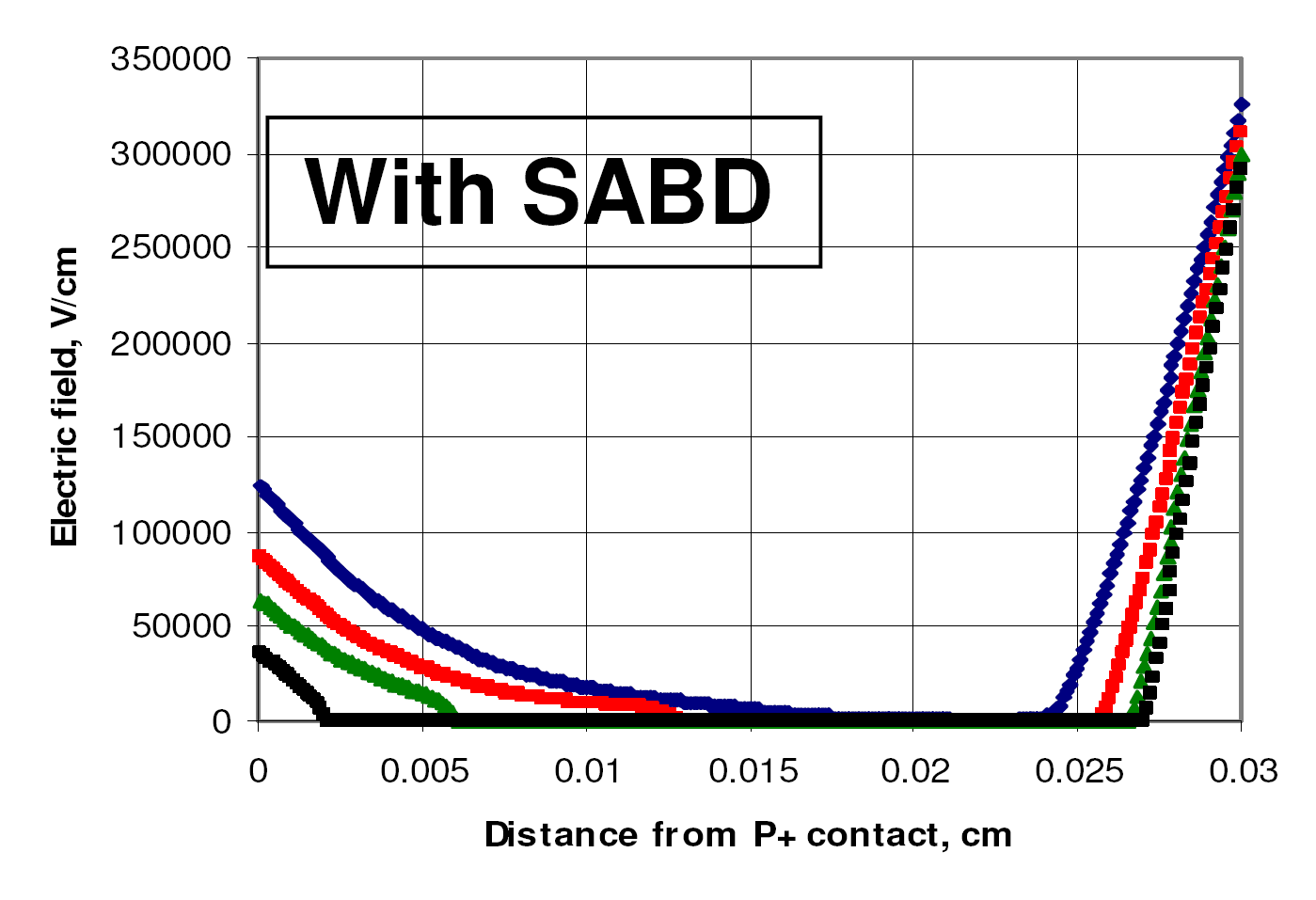}
\caption{\label{f5} \it Left:
the Lorentz shift dependence on the fluence for electrons (triangles ir-, diamonds red-laser) and
holes (rectangles). Right: the electric field distribution for bias voltages of 1500, 1000, 750 and 500 V, respectively, in a sensor irradiated with $10^{16}~n_{eq}/cm^2$, as calculated by
the PTI model including Soft Avalanche Breakdown (SABD), as shown by V. Eremin at the RD50 meeting \cite{rd50}. The increase in Lorentz shift for holes for higher fluences can be explained by the
reduced electric field near the collecting p+ electrodes, while the decrease for electrons
can be explained by the increased electric field near the n+  electrode, where charge multiplication
may additionally lead to charge sharing preferably in the opposite direction, thus causing ''negative''
Lorentz shifts.  }
\end{center}
\end{figure}

% shifts measurements
The averaged signal position $\bar{x}$ was computed from
either a Gaussian fit or from the center of gravity of the
pulse heights $p_i$ on neighboring strips $x_i$, i.e. $ \bar{x}
= {\sum p_i\cdot x_i}/{\sum p_i}$; both methods gave comparable
results.
The dependence of the Lorentz shifts on fluences for different sensors is summarized in
Fig.\,\ref{f5}.

% direct discussion
The surprising result is that for holes the Lorentz shift increases with fluence, while
for electrons it decreases, so at high fluences the Lorentz shift for holes is significantly
larger than for electrons in contrast to the shifts for non-irradiated sensors.
This dependence can be qualitatively explained by the expected redistribution of the electric field
into a double peak structure in highly irradiated sensors, as will be discussed in the next section.

 Although the shifts for the
non-irradiated sensor could only be measured to the breakdown
voltage of about 300\,V, the dependence on voltage has been
estimated  for zero fluence using the model
given in Ref.\,\cite{alg:2002}. These estimates for different voltages are shown
for zero fluence on the left hand side of Fig.\,\ref{f5}, which allows to see the dependence
of Lorentz shift versus fluence for a constant bias voltage.

% negative shifts
Interestingly, for electrons the Lorentz shift changes sign after a fluence of about 1.*10$^{15}$\,n$_{eq}$/cm$^2$.
The signal has still the same amplitude as for non-irradiated sensor, so
one still observes electrons, but shifted against the direction expected from the Lorentz force.
This shift was found to be hardly dependent on temperature, which  means it is not sensitive
to the drift mobility $\mu$. A possible qualitative explanation is discussed in the next section.


\section{Discussion and summary}
The observed difference in the electrons and holes shifts can be related to the structure of the E field in
irradiated sensors. After enough irradiation the bulk of the sensor is always p-type, since the irradiation induces acceptor-type defects at a high rate. So the pn-transition is always near the n-side of the diode, even if the initial material of the bulk was n-type. Type inversion from n-type to p-type depends on the initial doping concentration, but typically happens at a fluence of a few times $10^{13}$\,$n_{eq}/cm^2$. After a high fluence the detector cannot be fully depleted anymore for bias voltages up to 1000\,V, so the voltage drop mainly occurs on the undepleted layer near the n-side, which causes a high E-field near the n-side. Here soft avalanches can occur, which send holes back into the bulge, thus reducing the high electric field. This current stabilization and softening of the electric field  around the electrodes in heavily irradiated sensors was
discussed already  in 1977 \cite{eremin} and explains why irradiated sensors
can sustain much higher bias voltages than non-irradiated sensors. At high bias voltages the electric field
develops a significant field also near the p+ contacts,  as shown on the right hand side of Fig.\,\ref{f5}, which
explains, why one can see signals  from lasers shining from both sides of a heavily irradiated diode, i.e. one can study the movement of both, electrons and holes, in heavily irradiated sensors in spite
of the fact that the detector is operated below the depletion voltage.
On the  p+ side the electric field is much lower than on the  n+ side, so the holes  collected
on the p+ side will experience a higher Lorentz shift than electrons collected on the n+ side, because the Lorentz shift is proportional to B/E. This is in agreement with the data in Fig.\,\ref{f5}.

% negative shift
To explain the ''negative'' Lorentz shift for electrons at high fluences,
one has to explain simultaneously that: i) the collected charge is still high, although the sensor is
not fully depleted anymore; ii) the shift is in the opposite direction compared with the expectation from the
Lorentz force, as measured in the non-irradiated sensor; iii) the shift does not depend on temperature;
iv) the shift for a red  and infrared laser at intermediate fluences for which the
detector is just depleted at 1000\,V, have opposite signs, as is
visible in Fig.\,\ref{f5}.
%apparent from a comparison of Figs.\,\ref{f2} and \ref{f3}.
This means that a mechanism  different from the Lorentz shift expected from the
simple drift in ExB fields is operative.
% amplification

A hint for an explanation may come from the fact, that the same mechanism, which is responsible
for the high breakdown voltage after irradiation, namely the occurrence of charge multiplication
mentioned above, is also responsible for the negative Lorentz shift.
Indeed, a lot of evidence for charge multiplication has been collected recently,
as discussed at the RD50 meeting \cite{rd50}.
Most notably, one observes with an infrared laser shining from the edge of the detector
the occurrence of an increased current as soon as the electron cloud reaches the n+ contact and one observes a charge higher than expected from the induced ionization. Note that charge multiplication
is a common phenomenon in silicon devices; it
occurs when the electric field is typically above 15\,V/$\mu m$
and is e.g. responsible for the avalanches in silicon
photomultipliers or avalanche photo diodes. Fields above these values are indeed
expected in heavily irradiated sensor under high reverse bias
voltage, as shown on the right hand side of Fig.\,\ref{f5}.
% as shown by V. Eremin at the RD50 meeting \cite{rd50,eremin1}.
%

\begin{figure}
\begin{center}
\hspace {0mm}
\includegraphics [width=0.7\textwidth,clip] {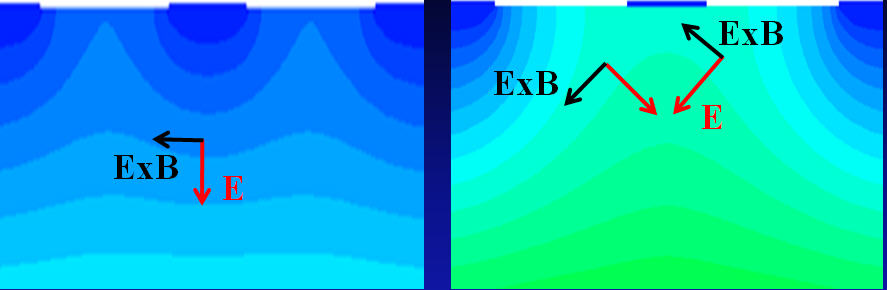}
\caption{\label{f6} \it Potential lines in a non-irradiated sensor (left) and irradiated sensor with a reduced electric
field on the central strip (right), which might be caused by charge multiplication (calculation with SYNOPSYS TCAD 2007  \cite{TCAD}). The electric field direction is perpendicular to the potential lines, as indicated by the arrows. The Lorentz force is in the direction of the ExB arrows.
The  horizontal electric field components from the neighboring strips can enhance
the charge multiplication in an asymmetric way, namely more charge is produced on the side, where the electrons are driven back to the strip by the Lorentz force (the right hand side of the central strip in the irradiated sensor).
This leads to asymmetric charge sharing,  thus causing the weighted charge distribution to shift in the direction opposite to the one expected from the Lorentz force.}
\end{center}
\end{figure}

How such a charge multiplication may lead to a Lorentz shift in the ''wrong'' direction
can be qualitatively understood as follows.
The reduced electric field on the strip with the highest charge multiplication leads to horizontal
electric field components from the neighboring strips, as indicated  on the right hand side of
Fig.\,\ref{f6}. These horizontal components in the reduced electric field
below the strip cause an asymmetry in the avalanche, because the Lorentz force, indicated
by the ExB vector on the right hand side (rhs) of Fig.\,\ref{f6}, drives electrons back to
the strip on the right hand side and away from the strip on the left hand side (lhs).
This means that the charge multiplication is stronger on the rhs than lhs,
thus providing a ''negative'' Lorentz shift by the charge sharing between the strips.
 For the red laser the electrons have to drift through
the whole sensor and the  Lorentz shift remains positive, as shown before in Fig.\,\ref{f2} on the rhs.
The charge multiplication on the n+ contact may shift it  to the opposite
direction again, thus explaining why the total shift is so low.
To obtain a quantitative understanding of the negative Lorentz shifts would require a detailed
simulation of a heavily irradiated sensor, which is difficult and beyond the scope of the present
paper.

% summary
In summary, the Lorentz shifts have been measured for holes and electrons in heavily irradiated sensors.
After heavy irradiation the sensors have a break down voltage well above 1000 V and at these high voltages
signals from both n+ and p+ strips can be observed in spite of the fact that the sensors are not depleted anymore. This can be explained by the well-known double peak model of the electric field in such sensors originating from charge multiplication at the n+ strips which reduces the electric field below the breakdown voltage by the so-called soft-avalanche breakdown mechanism. The double peak structure in the electric field leads to a low electric fields near the p+ strips where holes are collected and high electric fields near the n+ electrodes where electrons are collected. Consequently, electrons experience a lower Lorentz shift (proportional to B/E) than holes in heavily irradiated sensors in contrast to non-irradiated sensors, where the opposite is true.

\section{Acknowledgment}
We like to thank Gianluigi Casse for providing us with the Micron-sensors, which were produced in the framework of the RD50 collaboration.
Further we like to thank Jaako H\"ark\"onen for providing us with Magnetic-Chochalski-sensors, which were produced at the Helsinki Institute of Physics and Geoff Hall for providing us with the Premux chips.


\begin{thebibliography}{99}
\bibitem{smith} R.A. Smith,
Semiconductors,
{\em Cambridge Univ. Press\/},
1968.
%
 \bibitem{lb}
Landolt-B\"ornstein,
{\em Numerical Data and Functional Relationships in Science and Technology\/},
Group  III,
Vol. {\bf17a},
Springer Verlag,
Berlin,
1982.
\bibitem{JacoboniCanali:1977}
C. Jacoboni, C. Canali, G. Otiaviani and A. Albrrigi Quaranta,
A review of some charge transport properties of silicon,
{\em Solid-State Electronics \/}, Vol.{\bf 20}, pp.{\bf 77-89}, 1977. Pergamon Press.
\bibitem{eremin1}
V. Eremin, E. Verbitskaya and Z. Li,
% "The Origin of Double Peak Electric Field Distribution
%  in Heavily Irradiated Silicon Detectors", NIM A 476 (2002)  556.
{\em Nucl. Instr. and Meth. A\/} {\bf 476} (2002) 556.
\bibitem{jumbo}
F. Hornung, A. Rimikes, Th. Schneider,
High Magnetic Field facilities and Projects at the Forschungszentrum
Karlsruhe,
Internal Note, 1999.
%
\bibitem{roederer:1998} F. R\"oderer, Messung von
    Lorentz-Winkeln in Silizium-Detektoren, Diplomarbeit, Univ.
    of Karlsruhe, IEKP--KA/98--24 (german only).
%
\bibitem{heising:1999} S. Heising, Silicon detectors for high
    energy physics experiments at low temperatures and high
    magnetic fields, Ph. D. thesis. Univ. of Karlsruhe,
    IEKP--KA/99--26 (in german only).
%
\bibitem{hauler:2000} F. Hauler, Lorentzwinkelmessungen an
    bestrahlten Silizium-Streifendetektoren im
    Temperaturbereich T=77-300K, Diplomarbeit, Univ. of
    Karlsruhe, IEKP--KA/2000--12 (in german only).
%
\bibitem{osaka:2000} W. de Boer et al., Lorentz angle
    measurements in irradiated silicon detectors between 77K
    and 300K, {\em Nucl. Instr. and Meth. A\/} {\bf 461}
    (2001), 200--203; {\em Nucl. Instr. and Meth. A\/} {\bf
    478} (2002), 330-332;
%
\bibitem{Schneider:2009} M. Schneider, Lorentzwinkelmessungen
    in hochbestrahlten Siliziumstreifendetektoren, diploma
    thesis, Univ. of Karlsruhe, IEKP-KA/2009-14 (2009) (german
    only)
\bibitem{jones} L.L. Jones, PreMux128 Specification version
    2.3, Rutherford Internal Note, 1995.
\bibitem{alg:2002} Bartsch, V. et al., Nucl. Instrum. Meth. {\bf A497} (2003) 389,
http://arxiv.org/abs/physics/0204078v2
\bibitem{eremin} V. Eremin et. al.,
% "Scanning Transient Current
%    Study of the I-V Stabilization Phenomenon in Silicon
%    Detectors Irradiated by Fast Neutrons",
{\em Nucl. Instr. and Meth. A\/} {\bf 388} (1977) 350.
%
\bibitem{rd50} At the 15th Workshop of the RD50
    collaboration (http://rd50.web.cern.ch/rd50/) in November 2009 at CERN, Geneva,
    many new experiments supporting the charge multiplication by impact ionization were discussed, see
    contributions by V. Eremin, G. Kramberger, J. Lange and M. Milovanovic,  on tuesday, 17.11.2009, available at
    http://indico.cern.ch/conferenceOtherViews.py?view=cdsagenda\&confId=65918\#6

\bibitem{TCAD} http://www.synopsys.com/tools/tcad/Pages/default.aspx

\end{thebibliography}
\end{document}